\documentclass[conference]{IEEEtran}
\IEEEoverridecommandlockouts
\let\oldvec\vec
\usepackage{xcolor}
\usepackage{breqn}
\usepackage{relsize}
\usepackage{interval}
\usepackage{tabularx}
\usepackage{multirow}
\usepackage{amsmath,etoolbox}
\usepackage[ruled,vlined]{algorithm2e}
\usepackage{cite}
\usepackage{amsmath,amssymb,amsfonts}
\usepackage{algorithmic}
\usepackage{graphicx}
\usepackage{textcomp}
\usepackage{xcolor}
\usepackage{enumitem}
\setlist{nolistsep,leftmargin=.6cm}
\def\BibTeX{{\rm B\kern-.05em{\sc i\kern-.025em b}\kern-.08em
    T\kern-.1667em\lower.7ex\hbox{E}\kern-.125emX}}
\usepackage{amsmath, amssymb, bm, cite, epsfig, psfrag}
\usepackage{graphicx}
\usepackage{cuted}
\usepackage{soul} 
\usepackage[margin=0.7in]{geometry}
\usepackage{dblfloatfix}
\usepackage{array}
\usepackage{lipsum}
\newcolumntype{P}[1]{>{\centering\hspace{0pt}}p{#1}}
\newcolumntype{M}[1]{>{\centering\hspace{0pt}}m{#1}}
\newcolumntype{L}{>{\centering\arraybackslash}m{3cm}}
\usepackage[font=small]{caption}
\usepackage{epstopdf}	
\usepackage{longtable}
\usepackage{supertabular,booktabs}
\usepackage{bbm}
\usepackage{multirow}
\renewcommand{\arraystretch}{1.5}
\usepackage{subfigure}
\usepackage{etoolbox}
\usepackage{pbox}
\usepackage{fixltx2e}
\usepackage{tabu}	
\usepackage{filecontents}
\usepackage{enumerate}
\usepackage{textcomp}

\usepackage{colortbl}
\usepackage{fancyhdr}

\usepackage{bm}
\newtoggle{conference}
\togglefalse{conference} 
\def\argmin{\mathop{\mathrm{arg\,min}}}

\makeatletter
\newsavebox\myboxA
\newsavebox\myboxB
\newlength\mylenA
\newcommand*\xoverline[2][0.75]{%
    \sbox{\myboxA}{$\m@th#2$}%
    \setbox\myboxB\null
    \ht\myboxB=\ht\myboxA%
    \dp\myboxB=\dp\myboxA%
    \wd\myboxB=#1\wd\myboxA
    \sbox\myboxB{$\m@th\overline{\copy\myboxB}$}
    \setlength\mylenA{\the\wd\myboxA}
    \addtolength\mylenA{-\the\wd\myboxB}%
    \ifdim\wd\myboxB<\wd\myboxA%
       \rlap{\hskip 0.5\mylenA\usebox\myboxB}{\usebox\myboxA}%
    \else
        \hskip -0.5\mylenA\rlap{\usebox\myboxA}{\hskip 0.5\mylenA\usebox\myboxB}%
    \fi}

\begin{document}

\title{ Sparse  Activity Discovery in Energy Constrained Multi-Cluster IoT Networks Using Group Testing}

\author{\IEEEauthorblockN{Jyotish Robin, Elza Erkip}
\IEEEauthorblockA{\textit{Dept. of Electrical and Computer Engineering,}\\
\textit{Tandon School of Engineering, New York University, Brooklyn, NY, USA}}
}

\maketitle
\begin{abstract}
Current IoT networks are characterized by an ultra-high density of devices with different energy budget constraints, typically having sparse and sporadic activity patterns. Access points require an efficient strategy to identify the active devices for a timely allocation of resources to enable massive machine-type communication. Recently, group testing based approaches have been studied to handle sparse activity detection in massive random access problems. In this paper, a non-adaptive group testing strategy is proposed which can take into account the energy constraints on different sensor clusters.
A theoretical extension of the existing randomized group testing strategies to the case of multiple clusters is presented and the necessary constraints that the optimal sampling parameters should satisfy in order to improve the efficiency of group tests is established. The cases of fixed activity pattern where there is a fixed set of active sensors and random activity pattern where each sensor can be independently active with certain probability are examined.  The theoretical results are verified and validated by Monte-Carlo simulations. In massive wireless sensor networks comprising of devices  with different energy efficiencies, our proposed low-power-use mode of access can potentially extend the lifetime of battery powered sensors with finite energy budget.

\end{abstract}


\section{Introduction}~\label{sec:intro}
 Being a vital enabler for the digital metamorphosis in today's data driven world, IoT provides an ideal platform for a plethora of  applications in domains including, but not limited to smart cities, smart factories and smart homes. Ericsson \cite{Ericsson.} estimated that the cellular IoT growth will lead to 3.5 billion cellular IoT connections by 2023. The ITU-R workshop on IMT-2020 terrestrial radio interfaces \cite{ITUR} noted that the minimum requirement for connection density for evaluation in the mMTC usage scenario is 1,000,000 devices per km$^{2}$. These numbers are telltale of the accrescent demand for a unified connection fabric of things to cater to a multitude of smart sensing applications thereby making it viable for devices to autonomously function as a part of a smarter ecosystem.

 In comparison to the traditional cellular systems, IoT platforms come with several key differentiators that call in for an alternate perspective on system design. First of all, there is an ultra-high density of devices and hence the number of devices being managed by an access point (AP) can be quite large. Secondly, the activity pattern exhibited by the devices in the network is typically sparse (i.e., only a small fraction of the sensor population is active at a given time) and sporadic across the time domain due to the random nature of the events triggering the sensor activities. Furthermore, many IoT links need to support only low data rates in contrast to regular WiFi or cellular systems. Moreover, system energy usage is a critical aspect as many of the wireless sensors are meant to be low-power consumption devices. These differences have strong implications on how to facilitate active device discovery in IoT scenarios. For example, as pointed out in \cite{9060999}, the conventional coordinated multiple access schemes such as FDMA, TDMA, CDMA, SDMA and NOMA get extremely overloaded and are unable to perform the required coordination among sparse and sporadic users with  low latency requirements. On the other hand, ALOHA, the classical uncordinated multiple access strategy suffers from too many collisions while the more recent approach of Coded-Slotted ALOHA requires a large number of retransmissions thereby significantly reducing the efficiency.

Several novel schemes has been proposed in literature as candidate solutions to enable massive random access. In \cite{8323218}, an MMSE-based AMP algorithm is proposed for device activity detection exploiting the sparsity of the problem. The paper shows that in an asymptotic regime where the AP is equipped with a massive number of antennas, perfect activity detection is possible. Various compressive sensing based adaptive schemes are proposed in \cite{Ke_2020} by exploiting the sporadic traffic of massive connected devices and the virtual angular domain sparsity of massive MIMO channels.
Recently, there has been an emergence of group testing (GT) based methods for enabling sparse activity detection in massive random access scenarios. Specifically, in \cite{8262800}, a low-energy massive random access scheme for a single cluster of sensors based on non-adaptive group testing is studied where there is a global energy constraint which is translated to a constraint on the number of times a sensor is tested. In this work, the group testing codewords were based on the Kautz-Singleton construction \cite{1053689}. 

In this paper, we consider a multi-cluster scenario where each cluster is characterized by a different level of activity and energy budget constraint. Our aim is to efficiently identify all the active sensors while taking into account the energy budget constraints on each cluster. Using a GT strategy during the active device discovery phase, multiple active sensors are allowed to transmit simultaneously. At the AP, a simple energy detector is employed to detect the presence of energy. The number of transmissions each sensor makes during the active device discovery phase is restricted to accommodate for the energy budget constraints.  We rely on a randomized Bernoulli design based GT strategy \cite{6763117} rather than using explicit deterministic constructions.

The remainder of this paper is organized as follows: In section \ref{Sysmodel}, we describe the system model. Section \ref{sec:GT} introduces GT and provides an overview of the different  GT techniques relevant to the active device discovery problem. Section \ref{sec:proposedGT} describes the equivalence of  GT and active device discovery problem. Thereafter, we propose a modification to the randomized GT code based on Bernoulli design to account for the multi-cluster nature of the IoT environment. Subsequently, we derive constraints on optimal sampling parameters for the modified GT strategy which leads to efficient active device discovery  under random and fixed activity patterns. We also derive the optimal sampling parameters for a multi-cluster scenario with energy budget constraints on each cluster. Finally, we conclude our paper in Section \ref{con}.

\section{System model}~\label{Sysmodel}
Throughout the paper, we use the following notations and
definitions.
The set $\mathcal{U}$ =$\{s_{1},s_{2}, ...,s_{n}\}$ denotes the universe of sensors consisting of $n$ sensors. There are two possible states for each sensor: Active State $(\mathcal{AS})$ and Inactive State $(\mathcal{IS})$. We will reuse the notations $\mathcal{AS}$ and  $\mathcal{IS}$ to indicate the set of active and inactive sensors respectively. The meaning should be clear from the context. We assume that the sensors can be disjointly partitioned into $M$ clusters. The set of sensor clusters is represented  as $\mathcal{C}$ =$\{c_{1},c_{2}, ...,c_{M}\}$. The $M$ clusters have a deterministic number of sensors. i.e., $|c_{i}| = n_{i} $. $\mathcal{AS}_{j}$ and $\mathcal{IS}_{j}$ represents the set of active and inactive sensors in cluster $c_j$ respectively. Clustering of wireless sensors in IoT based on various metrics of interest has been well explored in the literature\cite{5636152}. In our model,  we assume that each cluster is characterized by a unique activity pattern as well as an energy budget (or energy efficiency) constraint.

We consider two  models for the activity pattern as follows:
\begin{itemize}
    \item Fixed activity: There is a fixed number of active sensors in each cluster with $\oldvec{\textbf{k}} = [k_{1},k_{2},...,k_{M}]$ denoting the number of active sensors in each of the $M$ clusters. Let $k=\sum_{i=1}^{M}k_i$.
    
    \item Random activity: In this model, each sensor becomes active independent of all the other sensors. $\oldvec {\textbf{p}_{a}} = [p_{1},p_{2},...,p_{M}]$ defines the probability of sensors being active in each of the $M$ clusters.
\end{itemize}
\color{black}

To model the energy budget constraint, we assume that the AP periodically assigns a $\beta_i$ value to cluster $c_i$ based on its relative energy efficiency. We assume that $\beta_i$'s are normalized such that $0 \leq \beta_i \leq 1$ and a higher $\beta_i$ corresponds to higher energy efficiency (or more available energy). The relative $\beta_i$ value essentially controls the number of transmissions  sensors in cluster $c_i$ perform during the active device discovery phase. Further details are described in Section \ref{sec:proposedGT}.
 
 To model the sparsity of the ``active set" we will be assuming that the number of active sensors ($k_i$) in the $i^{th}$ cluster scales sublinearly w.r.t the population size $(n_i)$. i.e.,  $k_i=\Theta(n_i^\alpha), \hspace{3pt} \text{where } \alpha \in(0,1)$. 

 In our approach, we will be using GT based strategy for energy-efficient active device discovery, the suitability of which will be described in the next sections.

\section{Overview on Group Testing}~\label{sec:GT}
Group testing can be viewed as a sparse inference problem where the objective is to identify a small number of ``defective" items from a large collection of items by performing tests on groups of items. In its standard form, each group test leads to a binary outcome where a \textbf{1} indicates the presence of at least one ``defective" in the group being tested and a \textbf{0} indicates the presence of zero defectives. The tests need to be devised such that the defective set of items can be recovered using the binary vector of test outcomes. Minimizing the number of tests is critical in many applications including the active device discovery problem we focus on.  Classifications of group testing models   relevant to our active device discovery problem are as below:

\vspace{0.2cm}

\noindent1) \textbf{Adaptive vs non-adaptive:}  In adaptive GT, the previous test results can be used to design the future tests. In non-adaptive setting, all group tests are designed independent of each other. 
    
\vspace{0.1cm}

\noindent2) \textbf{Small error vs zero error:} In small error setting, we aim to recover the defective set with high probability, i.e., the probability of error can be made arbitrarily small and vanishes asymptotically with the number of items. In zero error GT setting, we ensure that the defective set is certainly recovered.

A group testing matrix W is defined as a binary matrix formed by a  set of  $n$-coordinate column vectors (test vectors), $\mathbf{w}_{t} \in\{0,1\}^{n}$ where, $t \in\{1,2, \ldots, T\}$. i.e.,
\begin{equation}
    \mathbf{W}=\left[\mathbf{w}_{1}, \ldots, \mathbf{w}_{T}\right]=\left[\mathbf{x}_{1}, \ldots, \mathbf{x}_{n}\right]^{\intercal} \in \mathbb\{0,1\}^{n \times T}
    \label{GTmatrix}
\end{equation} where $T$ denotes the number of tests and $\mathbf{x}_i$ denote the $i^{th}$ row of the matrix corresponding to the test schedule for the $i^{th}$ item. Each column in a GT matrix is called a test vector.

Now, we will discuss some of the well known bounds that characterize the number of tests $T$ needed to recover a defective set of cardinality $k$ from a set of $n$ items.

\vspace{0.1cm}

\noindent 1) \textbf{Bounds on non-adaptive zero-error GT :} The non-adaptive zero-error GT usually relies on explicit design techniques for construction of \textit{disjunct} matrices \cite{1053689}. A binary group testing matrix, \textbf{W} as in (\ref{GTmatrix}) is $k$-disjunct if the Boolean sum of upto $k$ rows does not logically include any other row that is not a part of the summation. Bassalygo \cite{article} establishes that for the existence of a  $k$-disjunct $(n \times T)$-matrix, $T\geq min\{\binom{k+2}{2},n\}.$ 
This implies that, asymptotically, when $k=\Theta(n^{\alpha})$, where $ \alpha \geq \frac{1}{2}$, individual testing is the  optimal non-adaptive zero-error scheme. When  $\alpha < \frac{1}{2},$ the lower bound translates to $T\geq \Omega(k^{2})$. There are known  constructions of disjunct matrices discussed in literature \cite{doi:10.1142/1936},\cite{1053689} which are able to achieve $T=O(k^2\log n).$ For example, the Kautz-Singleton construction explained in \cite{1053689} requires $T=O(k^2\log^2_{k}n)$ which matches with $O(k^2\log n)$ in the sub-linear regime. 

\vspace{0.2cm}
   
\noindent 2)\textbf{ Bounds on non-adaptive  small-error  GT:}
    The non-adaptive small-error GT schemes of our interest are based on Bernoulli matrix design where all the entries of the group testing matrix \textbf{W} are independent samples from a Bernoulli random variable with an optimally designed parameter $q$. Achievability results  \cite{8926588} show that non-adaptive GT matrices with asymptotically vanishing probability of error can be constructed with $T=O(k\log n).$ There is a saving by a factor of $O(k)$ when we go from the zero-error to small-error setting.

\section{Group Testing based multi-cluster massive access}~\label{sec:proposedGT}
\vspace{-0.5cm}
\subsection{Group testing for sparse-Massive Random Access} 
Consider an $(n \times T)$- matrix $\textbf{W}$ in which the $i^{th}$ row is a binary signature of length $T$ designed for the $i^{th}$ sensor. In the active device discovery phase, each active sensor transmits its  binary signature (On-Off keying) in a time-synchronized manner over the $T$ probes. This potentially involves  a group of sensors transmitting at the same time if there are multiple active sensors with a $\textbf{1}$ at identical indices in their signatures. Note that in massive access, acquiring each device's channel state information (CSI) is impractical as it typically needs an overwhelming amount of pilot resources. Moreover, adaptively calibrating each device's channel is infeasible due to the massive device count \cite{9060999}. Thus, we assume that the decoder at the AP  makes a binary decision indicating the presence of energy in the received signal. This is essentially a non-coherent energy detector and hence does not require any CSI.  Let $\overrightarrow{\mathbf{y}}=\left(y_{t}\right) \in\{0,1\}^{T}$  indicate the results  vector. 

 \vspace{-.1cm}
\begin{equation}
y_{t}=\left\{\begin{array}{ll}
1 & \text { if energy detected }\left(\exists i \in \mathcal{A S} \text { with } \mathbf{w}_{t}(i)=1\right) \\
0 & \text { if no energy detected }\left(\forall i \in \mathcal{A S}, \mathbf{w}_{t}(i)=0\right)
\end{array}\right.
\end{equation}
 Given the matrix $\textbf{W}$ composed of binary signatures and results vector $(\oldvec{\textbf{y}}$), we need to identify (decode) which sensors are active in a computationally efficient manner. Clearly, this is equivalent to a group testing problem. The $t^{th}$ probe is a positive probe if $y_t=1$ and a negative probe if $y_t=0$. Thus, designing binary signatures for the $n$ sensors to detect the active sensors while minimizing the number of probes ($T$) is same as  designing a GT matrix and decoding the test results efficiently. Inan \textit{et al}. considered this model in \cite{8262800}.
 
 In our approach, we will be using a small error non-adaptive GT strategy. The non-adaptive nature allows us to probe multiple groups of sensors as dictated by the GT matrix simultaneously using other degrees of freedom (for eg: Frequency Division Multiplexing) thereby improving the latency performance. Furthermore, considering small-error instead of zero-error can reduce the number of probes (equivalently, the resource utilization) required during the active device discovery phase by a factor of $O(k)$ as we noted in Section \ref{sec:GT}. 
 
 In our study, we consider the  Combinatorial Orthogonal Matching Pursuit (COMP) decoding strategy, which is a practical and fast approach and is discussed thoroughly in the literature \cite{6763117,6120391} . COMP  classifies all participant sensors of a negative probe as inactive and all the remaining sensors as active. Note that COMP does not lead to any false negatives in comparison to the other non-adaptive decoding strategies such as Definitely Defective (DD), Sequential-COMP or Smallest Satisfying Set (SSS) \cite{8926588}. This is useful in an IoT environment which prevents misdetection of active sensors with critical information.
    
\subsection{Modified Bernoulli GT matrix for multi-cluster networks}
In the original Bernoulli design, each sensor is independently included in a probe (equivalently, group test) based on a ``global" sampling probability $q$ computed based on the sparsity of the problem. However, this scheme ignores the fact that there can be multiple clusters of sensors in the network with different energy budget constraints. In our paper, we bridge this gap by using a modified Bernoulli sampling strategy. Specifically, we include each sensor within cluster $c_{i}$  independently in a group test with probability $q_{i}$. Let $\oldvec q = [q_{1},q_{2},...,q_{M}]$ denote the vector of sampling probabilities. From a design point of view, we have the flexibility to choose $\oldvec q$ to optimize performance metrics of our interest while achieving successful decoding of sensor states without violating the energy constraints in place. We define \textit{success} as the event of inferring states of all sensors correctly. Thus, an \textit{error} occurs in COMP decoding if there are inactive sensors that are not part of any negative tests. We use the term \textit{shadowing} to denote the event in which a sensor is always tested along with at least one other active sensor in the entire probing session. Thus,
\begin{equation}
        \mathbb{P}(\mathrm{error})=\mathbb{P}\bigg(\bigcup_{s_{i} \in \mathcal{IS}}\{s_{i} \text { is shadowed }\}\bigg)
        \label{eq:COMP}
\end{equation} 

In the remainder of this section, we derive a constraint on the optimal $\oldvec{\textbf{q}}$ for COMP which minimizes the probability of error. Thereafter, we incorporate the energy budget constraints to derive the corresponding optimal sampling probabilities minimizing an upper bound on the probability of error.

\subsection{Optimal sampling parameters for COMP decoding}

\hspace{-.4cm} 1)\textbf{ Fixed Activity Pattern}:
In this case, there are M clusters  $[c_{1},c_{2}, ...,c_{M}]$ with a cardinality of $[n_{1},n_{2}, ...,n_{M}].$ Also, $ [k_{1},k_{2},...,k_{M}]$ represents the number of active sensors in each of the $M$ clusters. Chan \textit{et al.} \cite{6120391} derives the optimal sampling probability for a single cluster case using a union bound approach. We will be using a similar strategy along with reasonable approximations to derive the optimal sampling probabilities for a multi-cluster case. Using eq. (\ref{eq:COMP}), we can write,
\begin{subequations}
\begin{align}
 \hspace{0cm}\mathbb{P}(\mathrm{error})  \leq  \sum_{s_{i} \in \mathcal{IS}} \mathbb{P}( \{\text{sensor } s_{i} \text { is shadowed}\})\hspace{1.2cm}\label{eq:fixedDef1} &\\= \sum_{j=1}^{M}\sum_{s_{i} \in \mathcal{IS}_{j}} \bigg(1-q_{j}\Big(\prod_{r=1}^{M}(1-q_{r})^{k_{r}}\Big)\bigg)^{T}\hspace{.1cm}\label{eq:fixedDef2}\hspace{0.3cm}&\\ =\sum_{j=1}^{M}(n_{j}-k_{j}) \bigg(1-q_{j}\Big(\prod_{r=1}^{M}(1-q_{r})^{k_{r}}\Big)\bigg)^{T}\hspace{.05cm}\label{eq:fixedDef3}&\\ =: f(\oldvec{\textbf{q}})\hspace{5.3cm}
        \end{align}
\end{subequations}where, the term $\prod_{r=1}^{M}(1-q_{r})^{k_{r}}$ in (\ref{eq:fixedDef2}) represents the probability that none of the active sensors are selected. Equation (\ref{eq:fixedDef3}) uses the fact that the terms in the inner summation in (\ref{eq:fixedDef2}) remains constant within each cluster. Moreover, the number of inactive sensors in cluster $c_j$ is $(n_j-k_j).$

In order to identify the optimal sampling probabilities ($q_{j}^{*})$, we need to minimize $f(\oldvec{q})$. For the $(n_1,k_1)$- single cluster case  Chan \textit{et al.} considered in \cite{6120391}, (\ref{eq:fixedDef3}) reduces to $f(q_{1})=(n_1-k_1) \times \Big(1-q_1(1-q_1)^{k_{1}}\Big)^T$ which is minimized at $q_1 =\frac{1}{k_{1}+1}$. For ease of analysis of the multi-cluster case, let us define:
\begin{equation}
  \alpha(\oldvec{q})=\prod_{r=1}^{M}(1-q_{r})^{k_{r}}  
  \label{alpha}
\end{equation}
\begin{equation}
 \Tilde{n}_i =(n_i-k_i)
  \label{nitildedefn}
\end{equation}
Thus, (\ref{eq:fixedDef3}) implies
\vspace{-.1cm}
\begin{equation}
    f(\oldvec{q})=\sum_{j=1}^{M}\Tilde{n}_j \big(1-q_{j}\alpha\big)^{T}
    \label{fq}
\end{equation}
where, for brevity, we used $\alpha$ to denote $\alpha(\oldvec{q})$. 
In order to  minimize  $f(\oldvec{q})$, we use the method of Lagrange multipliers, with (\ref{alpha}) as an equality constraint  as follows:

\begin{align}
     \mathcal{L}(\oldvec{q})=\sum_{j=1}^{M}\Tilde{n}_j \big(1-q_{j}\alpha\big)^{T}+\lambda\left(\bigg(\prod_{r=1}^{M}(1-q_{r})^{k_{r}}\bigg) -\alpha\right)
\label{lagrange}
\end{align}
Taking the derivative of (\ref{lagrange}) $w.r.t$ $q_{i}, \forall i \in \{1,2,\ldots,M\}$, and equating it to zero, we get:
 
 \begin{equation}
\left(T \alpha \Tilde{n}_i\right)(1-q_i \alpha)^{T-1}+\lambda k_{i}(1-q_i)^{k_{i}-1}\bigg(\prod_{\substack{{r=1} \\ r\neq{i}\\
            }}^{M}(1-q_{r})^{k_{r}}\bigg)=0 \label{der1}
 \end{equation}
Taking the derivative of (\ref{lagrange}) $w.r.t$  $\alpha$ and equating to zero leads to:
 \begin{equation}
     \sum_{j=1}^{M}Tq_{j}\Tilde{n}_j \big(1-q_{j}\alpha\big)^{T-1}+\lambda=0
     \label{set1}
 \end{equation}
 Multiplying (\ref{der1})   by $\frac{q_i}{\alpha}$, we get:
\begin{equation}
T q_{i} \Tilde{n}_i\left(1-q_{i} \alpha\right)^{T-1}+\frac{\lambda q_{i} k_{i}}{\alpha}\left(1-q_{i}\right)^{k_{i}-1}\Big(\prod_{\substack{{r=1} \\ r\neq{i}\\
            }}^{M}(1-q_{r})^{k_{r}}\Big)=0
\label{eq1_simpli}
 \end{equation}
 Adding the set of equation in (\ref{eq1_simpli})  for all values of  $i$ in the set $ \{1,2,\ldots,M\}$ and using the value of $\alpha$ as in (\ref{alpha}), we get,
 \begin{equation}
      \sum_{i=1}^{M}Tq_{i}\Tilde{n}_i \big(1-q_{i}\alpha\big)^{T-1}
      +\lambda\sum_{i=1}^{M}\Bigg(\frac{ q_{i} k_{i}}{1-q_i}\Bigg)=0
 \label{set2}
 \end{equation}
Comparing (\ref{set1}) and (\ref{set2}), we can conclude that 
\begin{equation}
\sum_{i=1}^{M}\Bigg(\frac{ q_{i} k_{i}}{1-q_i}\Bigg)=1
\label{constrainfixed}
\end{equation}
Equation (\ref{constrainfixed}) is a constraint on the optimal sampling probability $q_i^*$.
 
 We incorporate the energy constraint as follows. Assume that we have a base sampling probability $q$ and each cluster uses a fraction of $q$ as its sampling probability. Specifically, define the sampling probability of $i^{th}$ cluster as
 \begin{equation}
     q_i=\beta_{i} \times q, \forall i \in \{1,2,\ldots,M\}.
     \label{qfrac}
 \end{equation}
One can think of $\beta_i$ as a predefined parameter for the $i^{th}$ cluster based on its energy budget constraint.
 Using (\ref{qfrac}) in (\ref{eq:fixedDef3}) leads to 
 \begin{align}
     \mathbb{P}(\mathrm{err})  \leq \sum_{j=1}^{M}(n_{j}-k_{j}) \exp (-Tq\beta_{j}\Big(\prod_{r=1}^{M}(1-q\beta_{r})^{k_{r}}\Big)) \nonumber&\\
     \leq \sum_{j=1}^{M}(n_{j}-k_{j}) \exp ( -Tq\beta_{j}e^{-q\sum_{r=1}^{M}\beta_{r}{k_{r}}}) \hspace{.6cm}
      \label{upbnd}
 \end{align}
 Note that since finding a closed form solution for $q_i^*$'s from (\ref{eq:fixedDef3}) seems infeasible for large values of $M$, we used the inequality $(1-x)^{N} \leq e^{-N x}$ when $0\leq x\leq 1$. The upper bound on probability of error indicated in (\ref{upbnd}) can be minimized to obtain the optimal $q$ as below:
 \begin{equation}
 q_{i}^*=\frac{\beta_i}{\beta_{1}k_{1}+\beta_{2}k_{2}+ \ldots +\beta_{M}k_{M}}
 \label{optQ}
 \end{equation}
 .

Note that for massive access scenarios where $n$ is sufficiently large, the derived $q_i^*$'s closely satisfy the constraint in (\ref{constrainfixed}). Also, in the simple case where there is only  one cluster, say $c_1$, characterized by the parameters $(n_1,k_1)$, we can see that (\ref{constrainfixed}) reduces to $q_1 =\frac{1}{k_1}$ which is in close agreement with the existing results in literature \cite{8926588}.

\begin{figure}   
	\centering
	\includegraphics[width=0.85\linewidth,height=65mm]{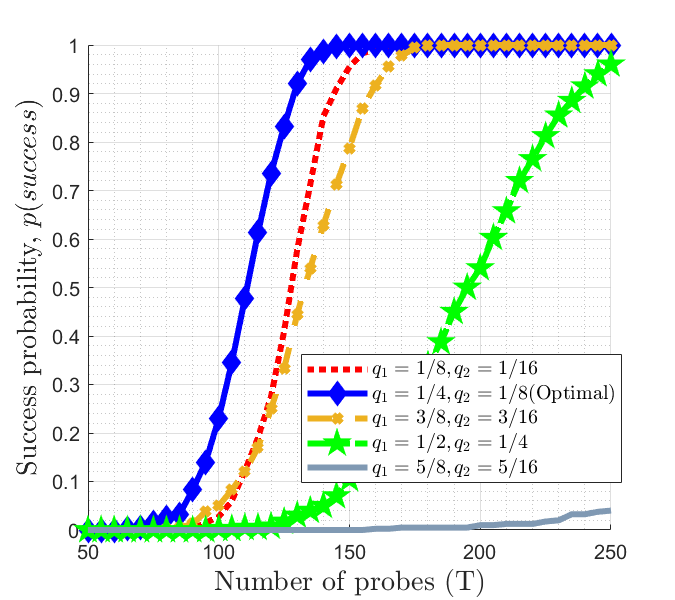}
	\setlength{\belowcaptionskip}{-10pt}
	\caption{Fixed activity pattern: Performance of the sampling probabilities derived for the 2-cluster case  with parameters $n_1=300, n_2=200, k_1=3, k_2=2,\beta_1=1,\beta_2 =0.5$ The derived optimal sampling probabilities are $q_1^{*}=\frac{1}{4}$ and $q_2^{*}=\frac{1}{8}$ (blue curve).}
	\label{fig:optimalSamp}
	\end{figure}

\begin{figure}   
	\centering
	\includegraphics[width=0.8\linewidth,,height=70mm]{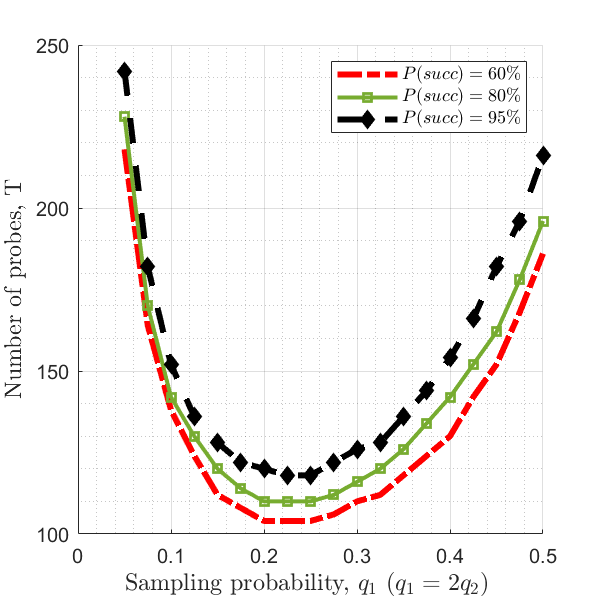}
	\setlength{\belowcaptionskip}{-14pt}
	\caption{Fixed activity pattern: The number of probes required versus the sampling probability to achieve a given $\mathbb{P}(Succ)$ in a 2-cluster case with parameters  $n_1=300, n_2=200, k_1=3, k_2=2,\beta_1=1,\beta_2 =0.5.$ The derived  sampling probabilities  $q_1^{*}=\frac{1}{4}$ and $q_2^{*}=\frac{1}{8}$ are close to the true minimum.}
	\label{fig:optimalSamp1}
	\end{figure}

As an example, consider a 2-cluster case with parameters $(n_1,k_1)=(300,3)$ and $(n_2,k_2)=(200,2)$. Assume that the base station assigned $\beta_1=1$ and $\beta_2=0.5$ indicating cluster-2 is having a   50\% lesser energy budget compared to cluster-1. The sampling probabilities are $q_1^*=\frac{1}{4}$ and $q_2^*=\frac{1}{8}$. It should be understood that though these sampling probabilities provide certain performance guarantees, they are obtained by minimizing an upper bound on probability of error and not the exact probability of error. We performed several empirical analysis to determine how close the parameter values we obtained from optimizing the union bound are to the best sampling probabilities. The utility of our derived sampling probabilities from a practical system design perspective is illustrated in Fig.\ref{fig:optimalSamp} and Fig.\ref{fig:optimalSamp1} . Clearly, the total number of probes required to achieve a given success probability using the derived  sampling probabilities  $q_1=\frac{1}{4}$ and $q_2=\frac{1}{8}$ are close to the true minimum leading to significant reduction in resource utilization.

	\vspace{-0.2cm}

\vspace{0.3cm}
\hspace{-0.2cm}2)\textbf{ Random Activity Pattern}:
Now, we proceed to analyze the case with random activity pattern. i.e., $\oldvec{\textbf{p}}_{a} = [p_{1},p_{2},...,p_{M}]$ defines the probability of activities for the sensors within each of the $M$ classes.

The inclusion of sensors in a test vector can be thought of as a collection of Bernoulli sampling processes. Each sensor in class $c_{i}$ is made part of the test group with probability $q_{i}$.
Since the sensor can be active with probability $p_{i}$, and inactive with probability $1-p_{i}$,  under independence assumption, an active sensor from $c_i$ is in a group test with probability 
$p_{i} q_{i}$ while an inactive sensor from $c_i$ is in a group test with probability $(1-p_{i})q_{i}.$

Using  Boole's inequality in (\ref{eq:COMP}), we can write the following:
\vspace{-0.5cm}
\begin{subequations}
\begin{align}
        \mathbb{P}(\mathrm{err})  \leq  \sum_{ \forall s_{i} } \mathbb{P}(\{s_{i} \text { is shadowed}\}\mid s_i \in \mathcal{IS})\mathbb{P}(s_i  \in \mathcal{IS})\hspace{5cm}\label{eq:comperr1}&\\= \mathlarger{\mathlarger{\sum}}_{j=1}^{M}\sum_{s_{i} \in c_{j}}\mathbb{P}(\{s_{i} \text { shadowed}\}\mid s_i \in \mathcal{IS}_{j})\mathbb{P}(s_i  \in \mathcal{IS}_{j})
        \label{eq:comperr2}\hspace{1.8in}&\\=\mathlarger{\mathlarger{\sum}}_{j=1}^{M}\Bigg(\sum_{s_{i} \in c_{j}}\bigg(1-   \frac{q_{j} \times\gamma(\oldvec{q})}{1-p_{j}q_{j}} \bigg) \Bigg) (1-p_j)\hspace{6.2cm}\label{eq:comperr3}&\\
        = \sum_{j=1}^{M}(n_j)(1-p_j)\times \Bigg(1-  \frac{q_{j} \times\gamma(\oldvec{q})}{1-p_{j}q_{j}}  \Bigg)^{T} \hspace{6.5cm}
        \label{eq:COMPbound}\end{align}
\end{subequations} 
where
\vspace{-.5cm}\begin{equation}
  \gamma(\oldvec{q})=\prod_{k=1}^{M}(1-p_kq_k)^{n_{k}} 
  \label{gamma}.
\end{equation}In (\ref{eq:comperr2}), $\mathcal{IS}_{j}$ denotes the set of inactive sensors in cluster $c_{j}$. Eqn.(\ref{eq:comperr3}) uses the fact that shadowing for an inactive sensor means that there are  no negative probes (tests with $y_{t} =0$) in which the sensor $s_{i}$ is a part of. The term $\frac{ \gamma(\oldvec{q})}{1-p_{j}q_{j}} $ denotes the joint probability that each of the remaining $n_{j}-1$ sensors in cluster $c_{j}$ as well as all the sensors from the remaining  $M-1$ clusters  are either inactive or not selected.

For brevity, we will use $\gamma$ to denote $\gamma(\oldvec{q})$. Similar to the fixed activity pattern, we proceed to minimize the upper bound in (\ref{eq:COMPbound})  by using the method of Lagrange multipliers with  (\ref{gamma}) as the equality constraint. The Lagrange function is given by

\begin{align}
   \mathcal{L}(\oldvec{q})=\sum_{j=1}^{M}n_j(1-p_j) \Bigg(1-\frac{ q_{j}}{(1-p_{j}q_{j})}\gamma \Bigg)^{T} \nonumber +\\\lambda\Big(\prod_{k=1}^{M}(1-p_kq_k)^{n_{k}} -\gamma\Big) . 
   \label{lagrange2}
\end{align}

Taking the derivative of (\ref{lagrange2}) w.r.t $\gamma$ and  $q_{i}, \forall i \in \{1,2,\ldots,M\}$,   leads to a system of equations which can be solved to obtain the following constraint on the optimal sampling probabilities ($q_i^*$'s).
 \begin{equation}
    \sum_{i=1}^{M} n_ip_iq_i=1
     \label{optq_rand}
 \end{equation}

 Clearly, for the  single cluster case with parameters $(n_1,p_1)$,the optimal sampling probability is $q_1^{*}=\frac{1}{n_1p_1}$. Also, if all the $M$ clusters are constrained to use  the same sampling probability, i.e., if $q_i^{*}=q^*, \forall i$, then $q^{*}=\frac{1}{\sum_{i=1}^{M}n_ip_i}.$
\vspace{.2cm}

To study the energy constrained case,  we invoke the constraint on sampling probabilities  defined in (\ref{qfrac}) and using the inequality $(1-x)^{N} \leq e^{-N x}$ when $0\leq x\leq 1$, (\ref{eq:COMPbound}) reduces to 
\begin{subequations}
\begin{align}
        \mathbb{P}(\mathrm{err})  \leq  \sum_{j=1}^{M}(n_j)(1-p_j) \exp (  \frac{-Tq\beta_{j} }{1-p_{j}q\beta_{j}}\prod_{k=1}^{M}(1-qp_k\beta_k)^{n_{k}}   ) \hspace{7.8cm}
        \label{eq:COMPbound1}&\\
       \leq  \sum_{j=1}^{M}(n_j)(1-p_j) \exp \Big(  -Tq\beta_{j} e^{-q\mathlarger{\sum}_{k=1}^{M}n_kp_k\beta_k}  \Big) \hspace{8.3cm} \label{upbnd1}\end{align}
\end{subequations} 

The error in this approximation will be small for a sparse- massive random access scenario once the number of probing instances ($T$) is sufficiently high. The upper bound on probability of error indicated in (\ref{upbnd1}) can be minimized to obtain the optimal $q$ as below:
 \begin{equation}
 q_{i}^*=\frac{\beta_i}{\beta_{1}n_{1}p_{1}+\beta_{2}n_{2}p_{2}+ \ldots +\beta_{M}n_{M}p_{M}}
 \label{optQ1}
 \end{equation}
 
\begin{figure}   
	\centering
	\includegraphics[width=0.85\linewidth,height=65mm]{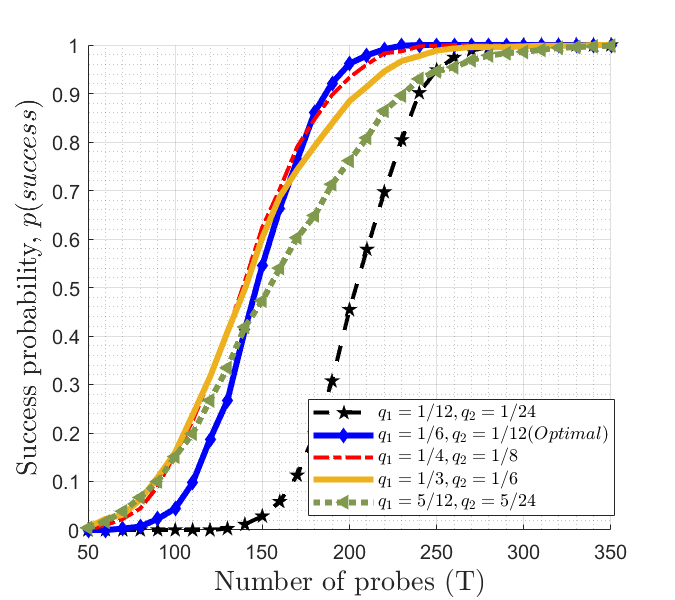}
	\setlength{\belowcaptionskip}{-14pt}
	\caption{Random activity pattern: Performance of the sampling probabilities derived for the 2-cluster case  with parameters $n_1=200, n_2=400, p_1=0.02, p_2=0.01,\beta_1=1, \beta_2 =0.5.$  The derived optimal  parameters are  $q_1^*=\frac{1}{6}$ and $q_2^*=\frac{1}{12}$ (blue curve).}
	\label{fig:prob_COMP}
	\end{figure}

\begin{figure}   
	\centering
	\includegraphics[width=0.8\linewidth,height=70mm]{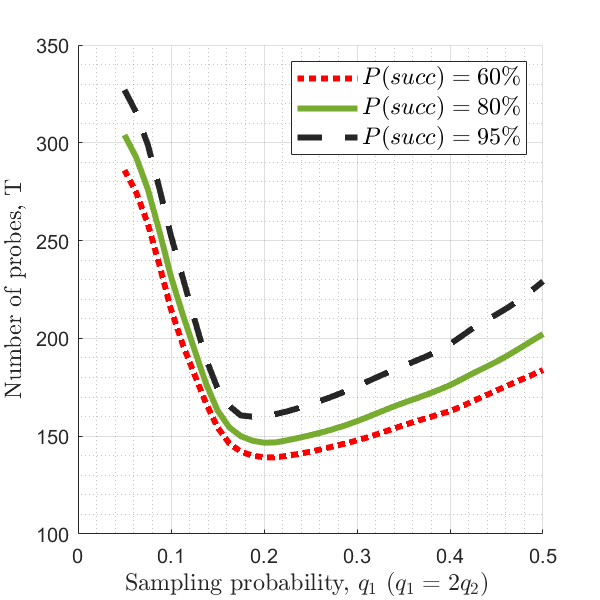}
	\setlength{\belowcaptionskip}{-10pt}
	\caption{Random activity pattern:  The number of probes required versus the sampling probability to achieve a given $\mathbb{P}(Succ)$ in a 2-cluster case with parameters  $n_1=200, n_2=400, p_1=0.02, p_2=0.01,\beta_1=1,\beta_2 =0.5.$ The derived optimal sampling probabilities  $q_1^{*}=\frac{1}{6}$ and $q_2^{*}=\frac{1}{12}$ are close to the true minimum.}
	\label{fig:prob_COMP2}
	\end{figure}

Simulation results demonstrating  the performance of the derived sampling parameters for a 2-cluster case is shown in Fig.\ref{fig:prob_COMP} and Fig.\ref{fig:prob_COMP2}. Clearly, once the number of probes reaches the usable regime where probability of success is considerably high,  derived sampling probabilities guarantee good performance. 

\section{Conclusion and Discussion} \label{con}
 In this paper, we have proposed and analyzed an active device discovery scheme for multi-cluster networks where each cluster of sensors is characterized by its own activity pattern and energy budget constraint. We have presented a modification to the original non-adaptive Bernoulli design based group testing strategy to account for the energy budget constraint on different clusters. Constraints on the optimal sampling probabilities ($\oldvec{\textbf{q}}$) have been derived for the fixed and random activity patterns respectively.
 
 In our approach, the energy budget (or, equivalently, energy efficiency) constraints for each cluster have been translated to a constraint on the corresponding sampling probabilities. Specifically, we assumed $q_i=\beta_i \times q$, where $\beta_i$ indicates the relative energy efficiency of the $i^{th}$ cluster and derived the optimal $q_i$'s. Through simulations, we have verified that our derived sampling probabilities can significantly reduce the resource utilization by bringing down the number of probes required to achieve a given probability of success.
 
 In a practical IoT setting, the energy-conserving group testing scheme we have proposed can extend the battery life of low-powered sensors without compromising its  network accessibility. This has significant impact in environments where batteries are difficult to replace or recharge periodically.
 
 Throughout this paper, we did not account for any  possible  sources of noise. Extensions to various noise models are part of future research.

\vspace{-0.3cm}
\bibliographystyle{IEEEtran}
\bibliography{SparseAct}

\begin{thebibliography}{10}
\providecommand{\url}[1]{#1}
\csname url@samestyle\endcsname
\providecommand{\newblock}{\relax}
\providecommand{\bibinfo}[2]{#2}
\providecommand{\BIBentrySTDinterwordspacing}{\spaceskip=0pt\relax}
\providecommand{\BIBentryALTinterwordstretchfactor}{4}
\providecommand{\BIBentryALTinterwordspacing}{\spaceskip=\fontdimen2\font plus
\BIBentryALTinterwordstretchfactor\fontdimen3\font minus
  \fontdimen4\font\relax}
\providecommand{\BIBforeignlanguage}[2]{{%
\expandafter\ifx\csname l@#1\endcsname\relax
\typeout{** WARNING: IEEEtran.bst: No hyphenation pattern has been}%
\typeout{** loaded for the language `#1'. Using the pattern for}%
\typeout{** the default language instead.}%
\else
\language=\csname l@#1\endcsname
\fi
#2}}
\providecommand{\BIBdecl}{\relax}
\BIBdecl

\bibitem{Ericsson.}
\BIBentryALTinterwordspacing
Ericsson mobility report 2017 nov, stockholm, sweden. [Online]. Available:
  \url{https://www.ericsson.com/assets/local/mobility-report/documents/2017/ericsson-mobility-report-november-2017.pdf}
\BIBentrySTDinterwordspacing

\bibitem{ITUR}
\BIBentryALTinterwordspacing
{ITU-R} {M.2410-0}: Minimum requirements related to technical performance for
  imt-2020 radio interface. [Online]. Available:
  \url{https://www.itu.int/dms_pub/itu-r/opb/rep/R-REP-M.2410-2017-PDF-E.pdf}
\BIBentrySTDinterwordspacing

\bibitem{9060999}
Y.~{Wu}, X.~{Gao}, S.~{Zhou}, W.~{Yang}, Y.~{Polyanskiy}, and G.~{Caire},
  ``Massive access for future wireless communication systems,'' \emph{IEEE
  Wireless Communications}, vol.~27, no.~4, pp. 148--156, 2020.

\bibitem{8323218}
L.~{Liu} and W.~{Yu}, ``Massive connectivity with massive mimo—part i: Device
  activity detection and channel estimation,'' \emph{IEEE Transactions on
  Signal Processing}, vol.~66, no.~11, pp. 2933--2946, 2018.

\bibitem{Ke_2020}
\BIBentryALTinterwordspacing
M.~Ke, Z.~Gao, Y.~Wu, X.~Gao, and R.~Schober, ``Compressive sensing-based
  adaptive active user detection and channel estimation: Massive access meets
  massive mimo,'' \emph{IEEE Transactions on Signal Processing}, vol.~68, p.
  764–779, 2020. [Online]. Available:
  \url{http://dx.doi.org/10.1109/TSP.2020.2967175}
\BIBentrySTDinterwordspacing

\bibitem{8262800}
H.~A. {Inan}, P.~{Kairouz}, and A.~{Ozgur}, ``Sparse group testing codes for
  low-energy massive random access,'' in \emph{2017 55th Annual Allerton
  Conference on Communication, Control, and Computing (Allerton)}, 2017, pp.
  658--665.

\bibitem{1053689}
W.~{Kautz} and R.~{Singleton}, ``Nonrandom binary superimposed codes,''
  \emph{IEEE Transactions on Information Theory}, vol.~10, no.~4, pp. 363--377,
  October 1964.

\bibitem{6763117}
C.~L. {Chan}, S.~{Jaggi}, V.~{Saligrama}, and S.~{Agnihotri}, ``Non-adaptive
  group testing: Explicit bounds and novel algorithms,'' \emph{IEEE
  Transactions on Information Theory}, vol.~60, no.~5, pp. 3019--3035, 2014.

\bibitem{5636152}
O.~{Boyinbode}, H.~{Le}, A.~{Mbogho}, M.~{Takizawa}, and R.~{Poliah}, ``A
  survey on clustering algorithms for wireless sensor networks,'' in \emph{2010
  13th International Conference on Network-Based Information Systems}, 2010,
  pp. 358--364.

\bibitem{article}
A.~Dyachkov and V.~Rykov, ``Survey of superimposed code theory.''
  \emph{Problems of Control and Information Theory}, vol.~12, pp. 229--242, 01
  1983.

\bibitem{doi:10.1142/1936}
\BIBentryALTinterwordspacing
D.-Z. Du and F.~K. Hwang, \emph{Combinatorial Group Testing and Its
  Applications}.\hskip 1em plus 0.5em minus 0.4em\relax WORLD SCIENTIFIC, 1993.
  [Online]. Available:
  \url{https://www.worldscientific.com/doi/abs/10.1142/1936}
\BIBentrySTDinterwordspacing

\bibitem{8926588}
M.~{Aldridge}, O.~{Johnson}, and J.~{Scarlett}, \emph{Group Testing: An
  Information Theory Perspective}, 2019.

\bibitem{6120391}
C.~L. {Chan}, P.~H. {Che}, S.~{Jaggi}, and V.~{Saligrama}, ``Non-adaptive
  probabilistic group testing with noisy measurements: Near-optimal bounds with
  efficient algorithms,'' in \emph{2011 49th Annual Allerton Conference on
  Communication, Control, and Computing (Allerton)}, Sep. 2011, pp. 1832--1839.

\end{thebibliography}
\end{document}